# South African night sky brightness during high aerosol epochs

H Winkler[1,3], F van Wyk[2] and F Marang[2]
[1] Department of Physics, University of Johannesburg, PO Box 524, 2006 Auckland Park, Johannesburg, South Africa
[2] South African Astronomical Observatory, PO Box 9, 7935 Observatory, Cape Town, South Africa

E-mail: hwinkler@uj.ac.za

**Abstract.** Sky conditions in the remote, dry north-western interior of South Africa are now the subject of considerable interest in view of the imminent construction of numerous solar power plants in this area. Furthermore, the part of this region in which the core of the SKA is to be located (which includes SALT) has been declared an Astronomical Advantage Zone, for which sky brightness monitoring will now be mandatory. In this project we seek to characterise the sky brightness profile under a variety of atmospheric conditions. Key factors are of course the lunar phase and altitude, but in addition the sky brightness is also significantly affected by the atmospheric aerosol loading, as that influences light beam scattering. In this paper we chose to investigate the sky characteristics soon after the Mount Pinatubo volcanic eruption in 1991, which resulted in huge ash masses reaching the stratosphere (where they affected solar irradiance for several years). We re-reduced photometric sky measurements from the South African Astronomical Observatory archives (and originally obtained by us) in different wavelengths and in a variety of directions. We use this data explore relationships between the aerosol loading and the sky brightness in a range of conditions, including several post-Pinatubo phases and during the passage of biomass burning induced haze and dust clouds. We use this data to explore the impact of our findings on the applicability of light scattering models and light scatterer properties.

## 1. Introduction
The quantification and spectral shape of sky brightness in the north western interior of South Africa has become of great importance for two quite different disciplines.

On the one hand, a vast region incorporating the Sutherland site of the South African Astronomical Observatory (SAAO) as well as the nucleus of the future Square Kilometer Array (SKA) has been declared an astronomical reserve, where light pollution is now restricted by law, and where sky conditions will have to be monitored regularly [1]. In addition, optimal use and scheduling of the SAAO Sutherland facilities (especially for the South African Large Telescope – SALT) requires the best possible knowledge of the sky illumination under all plausible conditions, e.g. aerosol loading and lunar phase and position.

Secondly, the broad geographic region adjacent to the astronomy reserve has been identified as the optimal site for a considerable number of solar power plants [2]. The energy yield of such plants is amongst other factors a function of the atmospheric transparency and the angular illuminance profile of the sky dome. In particular, the popular photovoltaic power generating method is (in addition to direct sunlight) also responsive to 'diffuse' irradiation (i.e. scattered skylight), and energy generation in this technology is furthermore critically dependent on the wavelength of the incoming photons.

---
[3] To whom any correspondence should be addressed.

While sky conditions in this region may during most days be considered as 'pristine', episodes of high aerosol concentration do occur periodically [3]. The bulk of these aerosols can be attributed to southern African biomass burning, local wind-induced dust production and occasional global-scale volcanic events.

The Mount Pinatubo volcanic eruption during the period immediately after 15 June 1991 resulted in the largest scale injection of aerosols into the stratosphere experienced in the $20^{th}$ century [4]. After several weeks the effects of stratospheric mixing resulted in these aerosols becoming spread around the globe. They first became detected over South Africa on 9 August 1991. Their concentration peaked soon thereafter, before gradually decreasing and reaching pre-eruption levels towards the end of 1994 [5].

**2. The data and its analysis**
In the course of normal astronomical photometric work, the brightness of an object of interest is determined by measuring the number of photons recorded from a patch of sky including this object ("star plus sky"), and subtracting from this value the brightness of a nearby and equally large empty patch of sky ("sky").

Tens of thousands of such sky background measurements have been made over the past decades by the authors, under a large variety of conditions, but such readings are normally used purely for calibration purposes, and have not been utilised in any way thereafter. It will be shown here how this archival data can be used to gain significant insights about sky brightness.

*2.1. Measurements*
The data was obtained with the 0.5 m telescope at the South African Astronomical Observatory in Sutherland. The light beam is first directed through a circular aperture at the focal point and then through a colour filter. Photons passing through these are finally recorded by means of a photomultiplier.

The circular aperture prevents light from nearby sources entering the detector, and hence limits the recorded light to that emanating from the circular patch of sky defined by the aperture.

Measurements were done through the U, B, V, R and I filters commonly used in astronomical work. Their transmittances as a function of wavelength [6] are illustrated in figure 1.

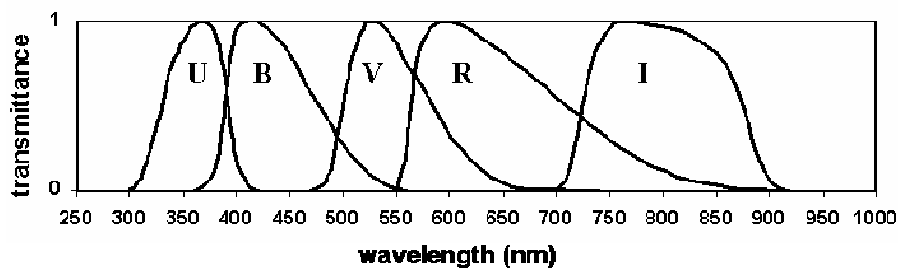

**Figure 1.** Relative transmittance of the colour filters used in this study.

*2.2. Selection of nights for analysis*
For this paper we have chosen to analyse some of the sky measurements made during the nights listed in table 1. These nights span a range of conditions encountered just prior to and soon after the

Pinatubo event. We also include an unrelated high-opacity episode to explore a different type of aerosol.

**Table 1.** Nights analysed in this study and the atmospheric conditions. The last five columns list the optical depth $\tau_\lambda$ at wavelengths representative of the colour filters used.

| Night | Aerosol | moon | $\tau_{(366\,nm)}$ | $\tau_{(438\,nm)}$ | $\tau_{(545\,nm)}$ | $\tau_{(641\,nm)}$ | $\tau_{(798\,nm)}$ |
|---|---|---|---|---|---|---|---|
| 26-27 Jun 1991 | low | yes | 0.497 | 0.249 | 0.138 | 0.092 | 0.065 |
| 24-25 Jul 1991 | low | yes | 0.497 | 0.249 | 0.138 | 0.092 | 0.065 |
| 9-10 Aug 1991 | low | no | 0.539 | 0.272 | 0.152 | 0.106 | 0.070 |
| 24-25 Oct 1991 | volcanic | yes | 0.640 | 0.345 | 0.212 | 0.180 | 0.134 |
| 20-21 Jan 1992 | volcanic | yes | 0.562 | 0.313 | 0.184 | 0.143 | 0.120 |
| 18-19 Feb 1992 | volcanic | yes | 0.562 | 0.313 | 0.203 | 0.157 | 0.129 |
| 10-11 Oct 1992 | volcanic/pyrogenic | yes | 0.589 | 0.332 | 0.212 | 0.180 | 0.138 |
| 21-22 Mar 2000 | aeolian or pyrogenic | yes | 0.585 | 0.295 | 0.157 | 0.115 | 0.074 |

In addition to describing the aerosol and lunar characteristics, table 1 also lists the optical depth $\tau$ in each of the filter bands measured during the night in question. The optical depth is defined as

$$\tau = -(\ln T)/X \quad . \tag{1}$$

$T$ refers to the atmospheric transmittance, and equals the fraction of the original light beam reaching ground level. The parameter $X$ is referred to the relative airmass, and designates the path length of a beam of light from the external radiation source through the atmosphere relative to the zenith atmospheric thickness. We adopt the following approximation for the airmass (which were also used in previous similar studies [7,8]), in terms of the zenith angle $\zeta$:

$$X = \left(1 - 0.96 \sin^2 \zeta\right)^{-0.5} \tag{2}$$

*2.3. Flux calibration*
For the chosen nights, the brightness measurements were calibrated through the observation of a large number of stars from the SAAO list of calibration standards [9]. During these nights the standard star observations were carried out at a wide range of zenith angles in order to minimise the errors in the optical depth value calculations.

**3. The theory of sky brightness at night time**
The luminance of the night sky is the combined contribution of a number of components:
- starlight (including the diffuse light of distant stars in the Milky Way and other galaxies).
- traces of evening or morning twilight.
- zodiacal light, which is caused by the reflection of sunlight by small solar system particles.
- airglow due to molecular transitions in the upper atmosphere, or other processes linked to atmospheric particles (auroral light, infrared thermal emission).
- light pollution in areas with ample artificial illumination sources.
- last, but not least, reflected moonlight, which can completely dominate the other contributions around the time of full moon [10].

The lunar brightness depends mainly on the angle Sun-Moon-Earth, but also on the Earth-lunar distance $D$. In particular, it needs to be considered that the lunar surface brightness is enhanced near full moon due to two optical effects referred to as shadow hiding and coherent backscatter [11]. We adopted a formulation derived from Noll et al [8],

$$\log B_\lambda(\omega, D) = \log B_\lambda(0, D_{av}) - 0.4 \times \left(0.026|\omega| + 4 \times 10^{-9} \omega^4\right) - 2\log(D/D_{av}) \quad (3)$$

where $B_\lambda$ is the lunar brightness in the filter band of wavelength $\lambda$, $\omega$ is the phase angle of the moon (in degrees, and equal to zero at full moon) and $D_{av}$ is the average distance of the moon.

The brightness of a patch of sky $\sigma$ due to the scattering of moonlight is a function not only of i) the lunar brightness, but also of ii) the lunar zenith angle and its corresponding lunar airmass $X_m$, iii) the sky patch zenith angle $X_s$, iv) the lunar and sky patch azimuths, v) the optical depth and vi) the appropriate atmospheric particle scattering function $f$ [7,8].

The scattering function sums contributions $f_R$ due to Rayleigh scattering by the smallest particles and $f_M$ due to Mie scattering by larger aerosols. The Rayleigh contribution may be assumed to take the form

$$f_R(\psi) \propto \left(C_R + \cos^2 \psi\right) \quad (4)$$

where $\psi$ is the angular distance between the moon and the observed sky patch (i.e. a direct function of the zenith and azimuth angles) and where $C_R$ is a constant known to be close to 1 ($C_R = 1.06$ was adopted in previous studies [7,8]). In these same studies the Mie contribution was assumed to take on a relationship of the form

$$f_M(\psi) \propto 10^{-\psi/C_M} \quad . \quad (5)$$

with $C_M$ taken to be 40°. The inherent inadequacies of the above approach are evident by the fact that proportionality constants ranging by a factor of 10 have been adopted for the $f_M$ expression in equation 5, and that the formulations do not consider any dependence on wavelength.

The true total scattering function can however also be determined empirically, by measuring the sky brightness $\sigma$ at a variety of wavelengths for a wide range of lunar and sky patch positions, and then utilising the relationship

$$f_\lambda(\psi) \propto \sigma_\lambda(\psi, X_s) \times \left[B_\lambda \exp(-\tau_\lambda X_m)(1 - \exp(-\tau_\lambda X_s))\right]^{-1}. \quad (6)$$

This procedure is adopted in this study. The locations of the Sun and Moon relative to the observing site can be determined to any desired accuracy. The freely available Solar and Moon Position Algorithm (SAMPA) used in this study calculates the solar and lunar coordinates to an accuracy of better than 10″ [12].

## 4. Results

After evaluating the measured sky brightness per unit solid angle (through division of the total flux by the area of the aperture), we used equation 6 to calculate the values of $f_\lambda$ for every sky measurement. When taking the logarithm of this quantity and plotting all values against the corresponding lunar-sky patch separation angle $\psi$ for each filter, we obtained a series of diagrams illustrated (for the U, V and I filters) in figure 2.

## 5. Discussion and conclusions

The graphs shown in figure 2 illustrate very clearly the shape and location of the scattering function $f(\psi)$, as well as the impact that wavelength and the presence of aerosols have on this. The increase in $f$ with growing optical depth can partly be explained as a consequence of a strengthening Mie component, which primarily affects those parts of the diagram where $\psi$ is small. That does however not account for the large discrepancies near $\psi \sim 90°$.

The visibly steeper nature of the plots for longer wavelength filters as opposed to what happens in the violet/UV diagram is also to be expected in view of the dominance of Rayleigh-scattered light in the blue part of the spectrum (even at night, blue photons dominate the radiation reflected off the atmosphere). The Mie fraction would thus be strongest at the longest wavelengths, explaining why the scattering function is considerably higher here for small $\psi$.

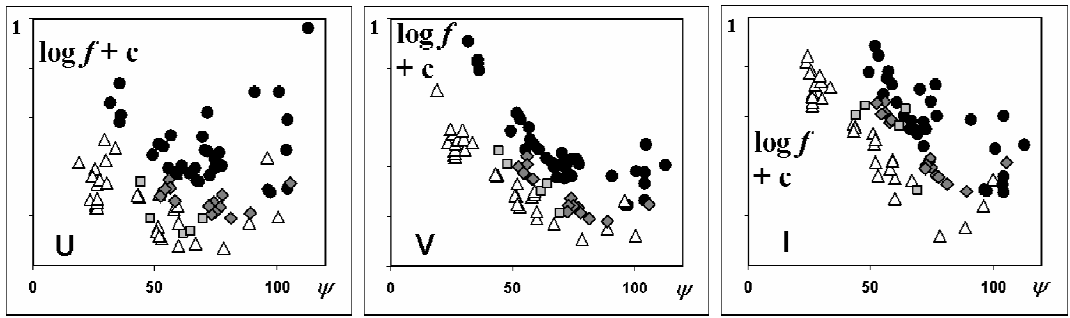

**Figure 2.** A comparison for the scattering function obtained in the U, V and I bands versus $\psi$ (in degrees) under the following conditions: i) pre-Pinatubo (white triangles); ii) early (Oct. 1991-Feb. 1992) post-Pinatubo (black circles); iii) late (Oct. 1992) post-Pinatubo (grey diamonds); iv) later (2000) dust event (grey squares).

Figure 2 also confirms the presence of the Rayleigh scattering function component. As predicted by the Rayleigh model in equation 4, $f$ reaches its minimum around $\psi = 90°$.

The discrepancy between the aerosol-free and aerosol-loaded curves is expected to be a consequence of the oversimplified assumptions that went into the derivation of equations by the previous studies [7]. A more sophisticated model formulation is required to adequately reproduce the interplay between attenuated moonlight and scattering events at various depths in the atmosphere. Such a model should also allow multiple scattering events.

Finally, the fact that the early post-Pinatubo plots are furthest removed from the 'low aerosol' curve (despite sharing similar opacity values with the plots associated with the later dates) indicates a probable dependence on the nature of the aerosols. It needs to be noted here that following volcanic eruptions the largest particles fall to the ground first due to their greater mass, while smaller particles stay airborne for considerably longer. This effect was also noted for the Pinatubo eruption [5]. It is projected that the planned systematic analysis of all available SAAO night sky data will be able to explore these and other questions in greater detail.

**Acknowledgments**
We thank successive Directors of the South African Astronomical Observatory for the generous allocation of observing time, and its numerous support staff for the good maintenance of the equipment. A special thanks to Di Cooper and Ramotholo Sefako for their assistance in the retrieval of archival data. Last, but not least, we thank Dave Kilkenny. Without his dedicated leadership of the SAAO photometry programme and his continual oversight of the quality control processes, this work would not have been possible.

**References**
[1] Jonas J L 2009 *Proc. Inst. Elec. Electron. Eng.* **97**, 8, 1522
[2] Fluri T P 2009 *Energy Policy* **37** 5075
[3] Formenti P, Winkler H, Fourie P, Piketh S, Makgopa B, Helas G and Andreae M O 2002 *Atmospheric Research* **62** 11

[4]  McCormick M P, Thomason L W and Trepte C R 1995 *Nature* **373** 399
[5]  Kilkenny D 1995 *The Observatory* **115** 25
[6]  Bessell M and Murphy S 2012 *Publ. Astron. Soc. Pacific* **124** 140
[7]  Krisciunas K and Schaefer B E 1991 *Publ. Astron. Soc. Pacific* **103** 1033
[8]  Noll S, Kausch W, Barden M, Jones A M, Szyska C, Kimeswenger S and Vinther J 2012 *Astron. & Astrophys.* **543** A92
[9]  Menzies J W, Cousins A W J, Banfield R and Laing J D 1989 *South African Astronomical Observatory Circulars* **13** 1
[10] Patat F 2003 *Astron. Astrophys.* **400** 1183
[11] Kieffer H H and Stone T C 2005 *Astron. J.* **129** 2887
[12] Reda I 2010 *Solar Eclipse Monitoring for Solar Energy Applications using the Solar and Moon Position Algorithms* NREL Report No. TP-3B0-47681